# A physical mechanism of solar corona heating

© 22. 04.2010    I. K. Mirzoeva

Space Research Institute, Russian Academy of Sciences,
ul. Profsoyuznaya 84/32, Moscow, 117997 Russia

e-mail: colombo2006@mail.ru

Time profiles of solar soft X-ray microflares and structure soft X-ray solar corona
thermal background are studied on RHESSI data. The observations of 2003 year are analyzed.
Decrease fluxe of solar soft X-ray microflares and thermal background of solar corona in
the X-ray range 2-15 kev are revealed. The new model of solar corona heating in based
on this new data are suggested.

УДК  523.985    **Механизм нагрева корональной плазмы**

© 22. 04.2010г.    И.К. Мирзоева

Институт космических исследований РАН
e-mail: colombo 2006@mail.ru

По данным проекта RHESSI исследованы временные профили вспышек малой мощности и структура излучения теплового фона солнечной короны в мягкой компоненте рентгеновского излучения Солнца за 2003г. Обнаружено падение интенсивности рентгеновского излучения солнечных событий и теплового фона короны в диапазоне энергий от 2 до 15 кэВ. На основе полученных результатов предложена новая модель нагрева солнечной короны.

На сегодняшний день рассматривается множество гипотез, объясняющих механизм нагрева корональной плазмы: нагрев корональной плазмы солнечными вспышками, нагрев ударными волнами, нагрев волнами конвективного шума. Существуют теории, согласно которым плазма солнечной короны нагревается постоянно циркулирующими фотосферными токами с последующей поставкой тепла в корону. Высказываются мнения и о многочисленных комбинациях всех механизмов. Мы не будем оспаривать ни одну из современных точек зрения. Единственное, на что хотелось бы обратить внимание – это серьезное замечание в адрес теории нагрева солнечной короны вспышками. В работах (Мирзоева И.К., Ликин О.Б., 2004, 2005); (Мирзоева И.К., 2006); (Christe S.D, 2007) было доказано существование предела в распределении солнечных вспышек по энерговыделениям, следовательно солнечных вспышек не бесконечное число, особенно крупных. Вспышки - это высокоэнергичные, но локальные явления и их энергии явно не достаточно для обеспечения и поддержания стабильного нагрева по всему объему короны.

На основе полученных экспериментальных данных предлагается механизм нагрева солнечной короны не связанный ни со вспышками, ни с ударными волнами, ни с токами.

Традиционно считается, что солнечная корона – так называемая открытая система относительно внешнего космического пространства, т.е. система в которой происходит свободная диссипация энергии во внешнее пространство. С поверхности Солнца в космос излучается весь спектр электромагнитных волн, уходят потоки заряженных частиц – солнечный ветер, гравитационные, звуковые, магнитогидродинамические волны, случаются выбросы корональной плазмы и др. Предположим, что в силу определенных причин (их физика потребует дополнительного изучения), корона – система не совсем открытая. Корональная плазма, на наш взгляд, имеет зоны непрозрачности для определенного спектра электромагнитных волн (рис.1). На рисунке такие зоны отмечены дугами со штрихами.

Согласно полученным экспериментальным данным в проектах RHESSI и INTERBALL, плазма солнечной короны может оказаться непрозрачной для потока мягкого рентгеновского излучения в узких полосках энергетического спектра в диапазоне от 2 до 15 кэВ. Явление падения интенсивности рентгеновского излучения, зарегистрированное по данным проектов RHESSI и INTERBALL, доказывает существование таких областей непрозрачности корональной плазмы. В результате взаимодействия постоянного потока электромагнитного излучения в энергетическом диапазоне 3-4 кэВ, энергия электромагнитной волны передается корональной плазме, вызывая стабильный нагрев до $2 \times 10^6$ К. При этом, энергия рентгеновского потока падает, что мы и наблюдаем в виде

явления падения интенсивности рентгеновского излучения в узкой полосе энергетического спектра (рис. 2)

Как известно, распространение электромагнитных волн в плазме (без учета влияния магнитного поля: в нашем случае магнитное поле короны не оказывает непосредственного влияния на нейтрально заряженный поток рентгеновского излучения) возможно только при частотах волны ω выше плазменной частоты $\omega_0$. Для случаев, при которых ω < $\omega_0$ волна отражается от границы плазмы. В нашем случае мы имеем частичное поглощение электромагнитной волны в энергетическом диапазоне от 2 до 5 кэВ корональной плазмой с последующим нагревом. Однако, простые расчеты по известным формулам показывают, что для волн данного диапазона мы имеем обратное соотношение частот: ω > $\omega_0$ для всех ω спектра от 2 до 5 кэВ.

Следовательно, объяснить данное явление, с помощью только простого взаимодействия электромагнитной волны с корональной плазмой не удается. Должны существовать дополнительные факторы, в силу которых в данном случае нарушается известное в физике плазмы соотношение частот.

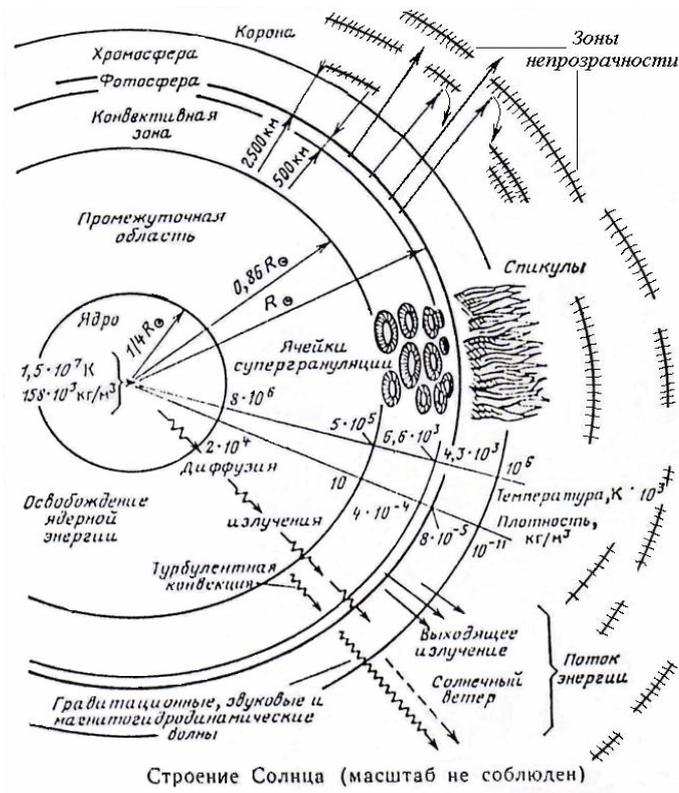

Рис. 1. Зоны непрозрачности корональной плазмы частично экранируют выходящее излучение в узких энергетических полосках электромагнитного спектра.

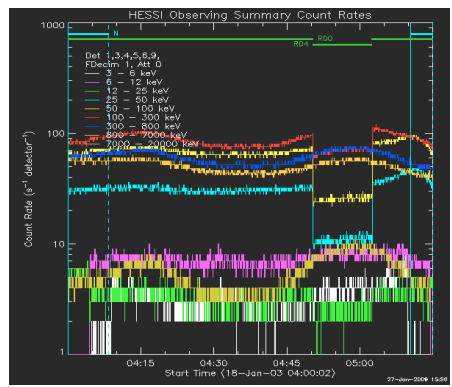

Рис.2. Данные, полученные в проекте RHESSI за 18.01.2003г. с 04:00 UT по 05:15 UT.